# LDPC Code Design for Transmission of Correlated Sources Across Noisy Channels Without CSIT

Arvind Yedla, Henry D. Pfister, and Krishna R. Narayanan
Department of Electrical and Computer Engineering
Texas A&M University
Email: {yarvind,hpfister,krn}@tamu.edu

*Abstract*—We consider the problem of transmitting correlated data after independent encoding to a central receiver through orthogonal channels. We assume that the channel state information is not known at the transmitter. The receiver has access to both the source correlation and the channel state information. We provide a generic framework for analyzing the performance of joint iterative decoding, using density evolution. Using differential evolution, we design punctured systematic LDPC codes to maximize the region of achievable channel conditions, with joint iterative decoding. The main contribution of this paper is to demonstrate that properly designed LDPC can perform well simultaneously over a wide range of channel parameters.

## I. PROBLEM SETUP

Consider the problem of transmitting the outputs of two discrete memoryless correlated sources, $(U_1, U_2)$, to a central receiver through two independent discrete memoryless channels with capacities $C_1$ and $C_2$, respectively. The system model is shown in Figure 1. We will assume that the channels belong to the same channel family, and that each channel can be parametrized by a single parameter $\alpha$ (e.g., the erasure probability for erasure channels). The two encoders are not allowed to communicate. Hence they must use independent encoding functions, which map $k$ input symbols ($\mathbf{U}_1$ and $\mathbf{U}_2$) to $n_1$ and $n_2$ output symbols ($\mathbf{X}_1$ and $\mathbf{X}_2$), respectively. The rates of the encoders are given by $R_1 = k/n_1$ and $R_2 = k/n_2$. The decoder receives $(\mathbf{Y}_1, \mathbf{Y}_2)$ and makes an estimate of $(\mathbf{U}_1, \mathbf{U}_2)$. This joint source-channel coding problem can be seen to be an instance of Slepian-Wolf coding [1] in the presence of a noisy channel.

over a large set of channel parameters. Therefore, by the symmetry of the problem, we assume that both the encoders use identical codes of rate $R$ (i.e., $R = k/n, n_1 = n_2 = n$). Reliable transmission over a channel pair $(\alpha_1, \alpha_2)$ is possible as long as the Slepian-Wolf conditions (1) are satisfied.

$$\frac{C_1(\alpha_1)}{R} \geq H(U_1|U_2)$$
$$\frac{C_2(\alpha_2)}{R} \geq H(U_2|U_1) \qquad (1)$$
$$\frac{C_1(\alpha_1)}{R} + \frac{C_2(\alpha_2)}{R} \geq H(U_1, U_2)$$

For a given pair of encoding functions of rate $R$ and a joint decoding algorithm, a pair of channel parameters $(\alpha_1, \alpha_2)$ is achievable if the encoder/decoder combination can achieve an arbitrarily low probability of error for limiting block-lengths ($k \to \infty$). We define the achievable channel parameter region (ACPR) as the set of all channel parameters which are achievable. Note that the ACPR is the set of all channel parameters for which successful recovery of the sources is possible for a fixed encoding rate pair $(R, R)$. We also define the Slepian-Wolf region as the set of all channel parameters $(\alpha_1, \alpha_2)$ for which (1) is satisfied. The Slepian-Wolf region for the erasure channel family is shown in Figure 2. We wish to find LDPC codes which result in large ACPRs, preferably the entire Slepian-Wolf region. Coding schemes which have large ACPRs are desirable because, for example, such a scheme can minimize the outage probability for non-ergodic channels.

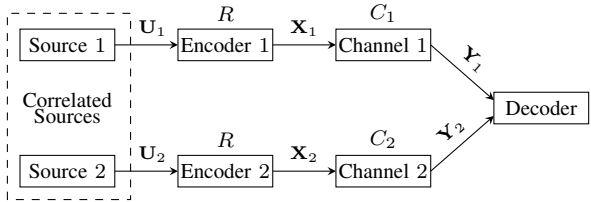

Figure 1. System Model

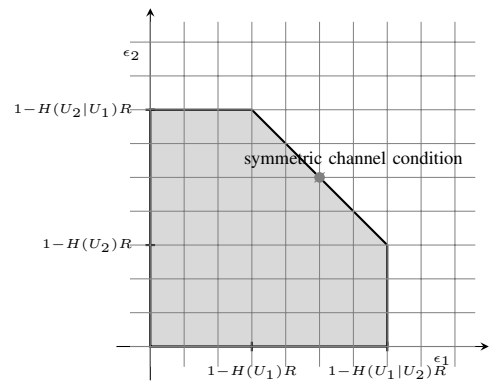

Figure 2. The Slepian-Wolf region for erasure channels, for a fixed rate pair $(R, R)$

The problem we consider is to design a graph based code, for which a joint iterative decoder can successfully decode

This work was funded by the Qatar National Research Foundation.

In this paper, we consider the following scenarios:
1) The channels are additive white Gaussian noise (AWGN) channels and the source correlation is modeled through a virtual correlation channel analogous to a binary symmetric channel (BSC).
2) The channels are erasure channels and the source correlation is modeled through erasures.

## II. BACKGROUND

### A. Prior Work

Recently this problem has attracted attention in the area of wireless sensor networks. It is a specific case of the sensor reachback problem [2]. Although separation between source and channel coding is known to be optimal for AWGN channels [2], it is desirable to take a joint source-channel coding approach (via direct channel coding and joint decoding at the receiver) [3]. When the channel parameters are known a priori at the transmitter, graph based coding schemes have been proposed [4], [5], which provide near optimal performance under joint iterative decoding. [3], [6] discuss the performance of concatenated LDGM codes under symmetric channel conditions. In [7], the authors study the performance of Turbo codes over varying channel conditions. Even when the channel parameters are known at the transmitter, it can be hard to design codes that perform well for symmetric channel conditions [8]. For the case of erasure correlated sources, our previous work gives a provable capacity achieving sequence of codes for the erasure channel under symmetric channel conditions [9].

However, in several practical situations, it is unrealistic for the transmitters to have a priori knowledge of the channel parameters. A single code now needs to perform well over the entire Slepian-Wolf region, making the problem of designing a good code even harder. In [7], [8], this problem is considered; however, the authors choose a code that performs well at one point on the Slepian-Wolf region and evaluate its performance for different channel parameters. As a result, the performance of the code is far from the optimal performance for some channel parameters. In this paper, we will show that by choosing an appropriate low density parity check (LDPC) code ensemble and optimizing it to perform well at several points on the Slepian-Wolf region, significantly better performance can be obtained.

### B. The Slepian-Wolf Region

Let $I_{\alpha_1}(X_1; Y_1)$ ($I_{\alpha_2}(X_2; Y_2)$) denote the mutual information between $X_1$ and $Y_1$ ($X_2$ and $Y_2$) when the underlying joint distribution is parametrized by $\alpha_1$ ($\alpha_2$). The following theorem shows the existence of codes which have large ACPRs.

*Theorem 1:* For a fixed pair of channel conditions $(\alpha_1, \alpha_2)$, which is not known at the transmitter, random coding with typical-set decoding at the receiver can achieve an average probability of error $\bar{P}_{e,\alpha_1,\alpha_2}$ (over all encoders with a rate $R$) bounded above by $2^{-n\mathsf{r}(\alpha_1,\alpha_2)}$, where

$$\mathsf{r}(\alpha_1, \alpha_2) = \min\big(I_{\alpha_1}(X_1; Y_1) - RH(U_1 \mid U_2),$$
$$I_{\alpha_2}(X_2; Y_2) - RH(U_2 \mid U_1),$$
$$I_{\alpha_1}(X_1; Y_1) + I_{\alpha_2}(X_2; Y_2) - RH(U_1, U_2)\big).$$

Hence, there exists an encoder for which the probability of error

$$P_{e,\alpha_1,\alpha_2} \leq 2^{-n\mathsf{r}(\alpha_1,\alpha_2)}.$$

*Proof:* The proof follows from standard random coding arguments. It is omitted here due to space constraints. ■

*Remark 1:* A simple application of Fano's inequality shows that any pair of channel parameters for which $\mathsf{r}(\alpha_1, \alpha_2) < 0$ are not achievable (the probability of error is strictly bounded away from zero). For binary memoryless symmetric (BMS) channels, the condition $\mathsf{r}(\alpha_1, \alpha_2) > 0$ translates to the conditions in (1). So, the conditions in (1) are both necessary and sufficient for transmission over BMS channels.

*Corollary 1:* For BMS channels, the achievable channel parameter region for a random code is a dense subset of the entire Slepian-Wolf region for limiting block-lengths.

*Proof:* Let S denote the Slepian-Wolf region and for $n \in \mathbb{N}$, let

$$\mathsf{A}_n = \{(q_i, r_i) \mid (q_i, r_i) \in \mathsf{S}, q_i, r_i \in \mathbb{Q}, i = 1, \cdots, n\}.$$

If $P_e^{(n)}$ denotes the random variable representing the probability of error of a randomly chosen encoding map with block-length $n$. Given $\delta > 0$, there exists $n \in \mathbb{N}$ such that $n^3 2^{-n\mathsf{r}(\alpha_1,\alpha_2)} < \delta$. Then,

$$\mathbb{P}(P_e^{(n)} < \delta \mid \alpha_1, \alpha_2) \geq 1 - \mathbb{P}(P_e^{(n)} \geq n^3 2^{-n\mathsf{r}(\alpha_1,\alpha_2)} \mid \alpha_1, \alpha_2)$$
$$\geq 1 - \frac{1}{n^3},$$

by the Markov inequality. We define the dominant exponent $\mathsf{r}_n = \min_{(\alpha_1,\alpha_2) \in \mathsf{A}_n} \mathsf{r}(\alpha_1, \alpha_2)$. Then we can upper bound the probability that a random code is *bad* for at-least one channel pair in $\mathsf{A}_n$ by a simple union bound, which gives us

$$\mathbb{P}(P_e^{(n)} \geq \delta) \leq \mathbb{P}(P_e^{(n)} \geq n^3 2^{-n\mathsf{r}_n}) \leq |A|_n \frac{1}{n^3} = \frac{1}{n}$$
$$\Rightarrow \lim_{n \to \infty} \mathbb{P}(P_e^{(n)} \geq \delta) = 0.$$

So, for limiting block-lengths, a single randomly chosen code can achieve a vanishing probability of error for a dense subset of the entire Slepian-Wolf region. ■

*Remark 2:* Corollary 1 also holds for random linear codes.

We conclude that, for a given rate pair $(R, R)$, a single encoder/decoder pair suffices to communicate the sources over all pairs of BMS channels in the Slepian-Wolf region. Thus, one can obtain optimal performance even without knowledge of $(\alpha_1, \alpha_2)$ at the transmitter. We refer to such encoder/decoder pairs as being *universal*. This means that random codes with typical-set decoding are universal for BMS channels.

Random coding with typical-set decoding is also universal with respect to channel types, since the performance of random coding depends only on channel capacity. In this paper, we are interested in universality with respect to channel parameters, for a particular channel type.

While random codes with typical-set decoding are universally good, this scheme is clearly impractical due to its large complexity. This motivates the search for low complexity encoding/decoding schemes which are universal. It was shown in [9] that codes that are good for the single user scenario need not be universal for the JSCC problem considered in this paper. It was conjectured in [8] that LDPC codes do not perform well for this problem[1]. To the contrary, we show that properly designed LDPC codes can have a large ACPR.

## III. SOURCE CORRELATION

In this section, we describe two possible correlation models between the sources. These models might appear restrictive, but will provide sufficient insight into the design of codes that perform well for arbitrary correlated sources. Our analysis in Section IV admits general correlation models.

### A. BSC Correlation

Consider a symmetric correlation model which can be defined in terms of a single parameter. For binary sources $U_1$ and $U_2$, this parameter is given by $\Pr(U_1 = U_2)$. It is useful to visualize this correlation by the presence of an auxiliary binary symmetric channel (BSC) with parameter $1 - p$ between the sources. In other words, $U_2$ is the output of a BSC with input $U_1$ i.e., $U_2 = U_1 + Z$. Here $Z$ is a Bernoulli-$(1-p)$ random variable and can be thought of as an *error*. Let $h_2(\cdot)$ denote the binary entropy function. Then, $H(U_1|U_2) = H(U_2|U_1) = h_2(p)$ and $H(U_1, U_2) = 1 + h_2(p)$.

This correlation model can be incorporated into the Tanner graph at the decoder (described in Section IV-B) as check nodes between the source bits, with a hidden node representing the auxiliary random variable $Z$ (which carries a constant log-likelihood ratio $\log \frac{1-p}{p}$) attached to the check node. For this scenario, the decoder does not require any side information i.e., it does not need to know the realization of the auxiliary random variable $Z$.

### B. Erasure Correlation

Another common way to model the source correlation is through erasures. Let $Z$ be a Bernoulli-$p$ random variable. The correlation between $U_1$ and $U_2$ is defined by

$$(U_1, U_2) = \begin{cases} \text{i.i.d. Bernoulli } \frac{1}{2} \text{ r.v.s, if } Z = 0 \\ \text{same Bernoulli } \frac{1}{2} \text{ r.v. } U \text{ , if } Z = 1 \end{cases}$$

We have $H(U_1|U_2) = H(U_2|U_1) = 1 - p$ and $H(U_1, U_2) = 2 - p$. This correlation model can be incorporated into the Tanner graph at the decoder with the presence or absence of a check node between the source bits depending on the auxiliary random variable $Z$. Note that the decoder requires the realization of the random variable $Z$ (for each of the source bits) as side information. This model can also be thought of as having two types of BSC correlation between the source bits, one with parameter 0 and one with parameter 1. The correlation parameter $p$ determines how many bits are correlated with parameter 1.

## IV. ANALYSIS

### A. Puncturing

It is shown in [10] that correlated codes are suboptimal when transmitting correlated sources over independent channels. The conditions in (1) implicitly assume the use of uncorrelated codes i.e., we require the average mutual information

[1]The authors consider only systematic LDPC codes

(over the code ensemble)

$$I(X_1; X_2) = 0.$$

This condition is clearly not satisfied when we use a systematic LDPC ensemble. This also explains the loss in performance of systematic LDPC codes when compared to Turbo codes, as shown in [8]. So, to ensure the independence of the transmitted codewords, we use LDPC ensembles which can be represented by a punctured systematic encoder.

### B. LDPC Codes

Assume that the sequences $\mathbf{U}_1$ and $\mathbf{U}_2$ are encoded using LDPC codes with a degree distribution pair $(\lambda, \rho)$, with a punctured systematic encoder. Let the fraction of punctured (systematic) bits be $\gamma$. Based on standard notation [11], we let $\lambda(x) = \sum_i \lambda_i x^{i-1}$ be the degree distribution (from an edge perspective) corresponding to the variable nodes and $\rho(x) = \sum_i \rho_i x^{i-1}$ be the degree distribution (from an edge perspective) of the parity-check nodes in the decoding graph. The coefficient $\lambda_i$ (resp. $\rho_i$) gives the fraction of edges that connect to the variable nodes (resp. parity-check nodes) of degree $i$. Likewise, $L_i$ (resp. $R_i$) is the fraction of variable (resp. parity-check) nodes with degree $i$. Also, let $\mathcal{V} = \{i \,|\, \lambda_i \neq 0\}$ and $\mathcal{P} = \{i \,|\, \rho_i \neq 0\}$ be the support sets of the variable and parity-check degree distributions respectively. The rate pair of the two codes after puncturing is $(R, R)$, where

$$R = R(\lambda, \rho) = \frac{1}{1-\gamma}\left(1 - \frac{\int_0^1 \rho(x)\,\mathrm{d}x}{\int_0^1 \lambda(x)\,\mathrm{d}x}\right). \quad (2)$$

The Tanner graph [11] for the joint decoder is shown in Figure 3. Code 1 corresponds to the bottom half of the graph,

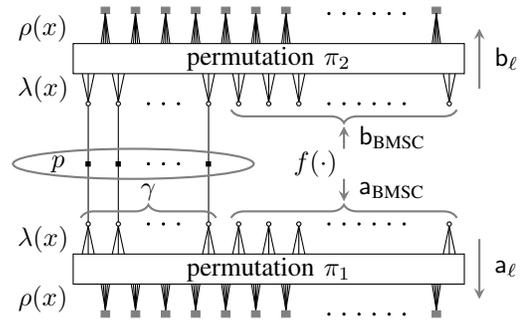

Figure 3. Tanner Graph of an LDPC Code with source correlation

code 2 corresponds to the top half and both the codes are connected by correlation nodes at the punctured bits. Let $\mathsf{a}_\ell$ and $\mathsf{b}_\ell$ denote the density[2] of the messages emanating from the variable nodes at iteration $\ell$, corresponding to codes 1 and 2. The density evolution equations [11] can be written as follows

$$\begin{aligned}\mathsf{a}_{\ell+1} &= \left[\gamma f\Big(L\left(\rho(\mathsf{b}_\ell)\right)\Big) + (1-\gamma)\mathsf{a}_{\text{BMSC}}\right] \circledast \lambda(\rho(\mathsf{a}_\ell)) \\ \mathsf{b}_{\ell+1} &= \left[\gamma f\Big(L\left(\rho(\mathsf{a}_\ell)\right)\Big) + (1-\gamma)\mathsf{b}_{\text{BMSC}}\right] \circledast \lambda(\rho(\mathsf{b}_\ell)),\end{aligned} \quad (3)$$

[2]Assuming that the transmission alphabet is $\{\pm 1\}$, the densities are conditioned on the transmission of the all one codeword (the all zero codeword assumption).

where $\lambda(\mathsf{a}) = \sum_i \lambda_i \mathsf{a}^{\circledast(i-1)}$, $L(\mathsf{a}) = \sum_i i \mathsf{a}^{\circledast(i-1)}$, $\rho(\mathsf{a}) = \sum_i \rho_i \mathsf{a}^{\boxplus(i-1)}$, $\mathsf{a}_{\text{BMSC}}$ and $\mathsf{b}_{\text{BMSC}}$ are the densities of the likelihood ratios received from the channel. The function $f$ at the correlation nodes depends on the equivalent channel corresponding to the correlation model, as described in [12]. For example, in the case of a BSC correlation with probability $p$, we introduce a parity-check at the correlation nodes which evaluates to a Bernoulli-$p$ random variable i.e., $f(\mathsf{a}) = \mathsf{a}_{\text{BSC}(p)} \boxplus \mathsf{a}$. For a BEC correlation with probability $p$, there is a parity-check at the correlation node with probability $p$ and with probability $1-p$ there is no parity-check, so $f(\mathsf{a}) = (1-p) + p\mathsf{a}$.

Using the error functional [11]

$$\mathfrak{E}(\mathsf{a}) = \int_{-\infty}^{0^-} \mathsf{a}(x)\,dx + \frac{1}{2}\int_{0^-}^{0^+} \mathsf{a}(x)\,dx,$$

the residual error probability at iteration $\ell$, $(e_1^\ell, e_2^\ell)$, is computed as

$$e_1^\ell = \mathfrak{E}\left(\left[\gamma f\left(L(\rho(\mathsf{b}_\ell))\right) + (1-\gamma)\mathsf{a}_{\text{BMSC}}\right] \circledast L(\rho(\mathsf{a}_\ell))\right)$$
$$e_2^\ell = \mathfrak{E}\left(\left[\gamma f\left(L(\rho(\mathsf{a}_\ell))\right) + (1-\gamma)\mathsf{b}_{\text{BMSC}}\right] \circledast L(\rho(\mathsf{b}_\ell))\right).$$

For two residual error probabilities $(e_1, e_2)$ and $(\tilde{e}_1, \tilde{e}_2)$, we define $(e_1, e_2) \preceq (\tilde{e}_1, \tilde{e}_2)$ iff $e_1 \leq \tilde{e}_1$ and $e_2 \leq \tilde{e}_2$.

### C. Differential Evolution

Throughout this section, we use $x$ to denote an element of $\mathbb{R}^n$ and $x_i$ to denote its $i$th component. The correlation parameter $p$ is fixed and the set of variable and parity-check degrees ($\mathcal{V}$ and $\mathcal{P}$) over which we need to optimize the degree profiles $\lambda$ and $\rho$ are assumed to be known. We design LDPC codes for this scenario using differential evolution [13], for a design rate $R_d$. Differential evolution has been shown to provide good results for designing LDPC codes over single user channels [14].

In an $n$-dimensional search space, a fixed number of vectors are randomly initialized and then evolved over time, exploring the search space, to locate the minima of the objective function. Let

$$\Delta^{n-1} = \left\{ x \in \mathbb{R}^n \,\bigg|\, \sum_{i=1}^n x_i = 1, x_i \geq 0, i = 1, \cdots, n \right\}$$

denote the unit simplex and $n_v = |\mathcal{V}|$, $n_p = |\mathcal{P}|$. Then, the search space for all variable (check) degree profiles is $\Delta^{n_v-1}$ ($\Delta^{n_p-1}$). The optimization is performed over the search space $\mathcal{S} = \Delta^{n_v-1} \times \Delta^{n_p-1}$, with parameter vectors $x = [x_\lambda, x_\rho]^3$, where $x_\lambda \in \Delta^{n_v-1}, x_\rho \in \Delta^{n_p-1}$.

In our optimization procedure, we expand the search space to $\mathcal{S}' = \{x \in \mathbb{R}^{n_v + n_p}, \sum_i (x_\lambda)_i = 1, \sum_i (x_\rho)_i = 1\}$, for simplicity in the crossover stage.

For the optimization to work well, differential evolution requires an initial population of trial vectors which are spread out uniformly across the search space. We generate an initial population of trial degree distributions by uniformly sampling the degree distributions from the unit simplex. To obtain a sample $x$ uniformly from $\Delta^{n-1}$, we generate uniform random variables $u_i \sim U[0,1], i = 1, 2, \cdots, n-1$. Define $u_0 = 0, u_n = 1$ and let $\pi_u$ be the permutation that sorts $(u_i)$ in ascending order i.e., if $i \leq j$, then $u_{\pi_u(i)} \leq u_{\pi_u(j)}$. For $i = 1, \cdots, n$, define $x_i = u_{\pi_u(i)} - u_{\pi_u(i-1)}$, and $x = (x_i)$. Then $x$ has a uniform distribution over $\Delta^{n-1}$.

---

[3]$(x_\lambda, \mathcal{V})$ and $(x_\rho, \mathcal{P})$ correspond to the variable and parity node degree profiles respectively.

Let $\mathcal{C}$ be a finite subset of channel parameters $(\alpha_1, \alpha_2)$ that correspond to the sum rate constraint of the Slepian-Wolf conditions for a design rate $R_d$. Let $\Gamma : \mathcal{S}' \times \mathcal{C} \to [0,1] \times [0,1]$, $(x, \alpha_1, \alpha_2) \mapsto (e_1, e_2)$ be the function that gives the residual error probability[4] (using joint density evolution as described in Section IV-B) for each decoder, for a pair of codes with degree distribution $x$ (i.e., $(x_\lambda, x_\rho)$), when transmitted over channels with parameters $(\alpha_1, \alpha_2)$. We use discretized density evolution [15] (with 9 bit linear quantization over a likelihood ratio range $[-20, 20]$).

For our design, we want the code to achieve an arbitrarily low probability of error on $\mathcal{C}$ and we want the rate of the code ($R(x)$) to be as close to the design rate ($R_d$) as possible. So, we define the cost function,

$$\mathcal{F}(x) = a \cdot \left( \sum_{(\alpha_1,\alpha_2) \in \mathcal{C}} \left(1 - \mathbb{1}_{\{(\alpha_1,\alpha_2) | \Gamma(x,\alpha_1,\alpha_2) \preceq (\tau,\tau)\}}\right) \right) + b \cdot (R_d - R(x)),$$

if $x \in \mathcal{S}$ and $\mathcal{F}(x) = \infty$, if $x \in \mathcal{S}' \setminus \mathcal{S}$. The rate of the code $R(x) = R(x_\lambda, x_\rho)$ is computed as in (2). The constants $a$ and $b$ are chosen through trial and error. The parameters chosen for the designs considered in this paper are $\tau = 10^{-5}, a = 10$ and $b = 30$. The optimization is then setup as

$$\min_{x \in \mathcal{S}'} \mathcal{F}(x).$$

We use a variant of differential evolution, with the following mutation and recombination scheme [14]. Parameter vectors $x_1, x_2, x_3$ and $x_4$ are selected at random from the current population. Let $x_b$ denote the best member of the current population. A trial vector $x$ is generated as

$$x = x_b + 0.5 \cdot (x_1 - x_2 + x_3 - x_4).$$

## V. RESULTS AND CONCLUDING REMARKS

We showed that the Slepian-Wolf conditions are necessary and sufficient for communication of correlated sources through independent BMS channels, without channel state information at the transmitter. This implies that a single random code is sufficient to communicate with vanishing probability of error, for the entire Slepian-Wolf region.

We designed punctured systematic LDPC codes for the scenarios described in Section III. The design was performed to maximize the ACPR, in contrast to previous work. The results are given below.

---

[4]We set the maximum number of iterations to 100 for all the designs considered in this paper. Density evolution is stopped when the maximum number of iterations are reached or the difference in the residual error probability between successive iterations is less than $10^{-8}$.

## A. BSC Correlation

We optimized the degree profiles for BSC correlated sources using the procedure described in Section IV-C, for transmission over the AWGN channel. The source correlation parameter was $p = 0.9$ and the optimization was performed for a design rate $R_d = 0.5$ after puncturing, resulting in the following degree profiles

$$\lambda(x) = 0.14895x + 0.3963x^2 + 0.13192x^3 + 0.18799x^6$$
$$+ 0.0050071x^8 + 0.064285x^{14} + 0.065546x^{19},$$
$$\rho(x) = 0.64719x^3 + 0.13662x^5 + 0.14429x^{14} + 0.071903x^{24}.$$

The ACPR for this code is shown in Figure 4. Also shown in the figure is the Slepian-Wolf region for the rate pair $(0.282, 0.282)$, which has a transmission rate pair $(0.423, 0.423)$ after puncturing. These results show that ensembles optimized using differential evolution almost achieve the entire Slepian-Wolf region. For a rate pair of $(0.423, 0.423)$, the best (theoretical) achievable signal to noise ratio (SNR) for symmetric channel conditions is $-2.68$ dB. The optimized code can perform well, for symmetric channel conditions, up-to an SNR of $-1.82$ dB, which is at a gap of $0.86$ dB from the theoretical limit.

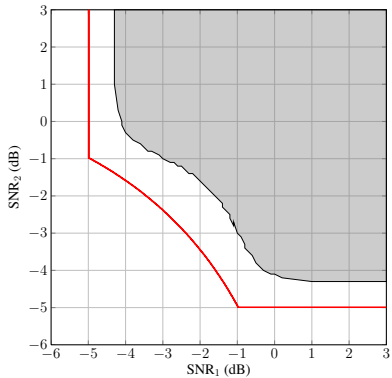

Figure 4. ACPR (Density Evolution threshold) of an optimized (AWGN channel) LDPC Code of rate 0.282

## B. Erasure Correlation

We use the procedure described in Section IV-C to design LDPC codes for the erasure channel, when the sources are correlated with an erasure probability of $p = 0.5$. The optimization was performed for a design rate of $R_d = 0.57$ after puncturing, resulting in the following degree profiles

$$\lambda(x) = 0.3633x + 0.2834x^2 + 0.2315x^6 + 0.1217x^{19},$$
$$\rho(x) = 0.531776x^3 + 0.468224x^5.$$

The ACPR for this code is shown in Figure 5. Also shown in the figure is the Slepian-Wolf region for the rate pair $(0.3308, 0.3308)$ which has a transmission rate pair $(0.4962, 0.4962)$ after puncturing. These results show that ensembles optimized using differential evolution almost achieve the entire Slepian-Wolf region.

## C. Future Work

The main motivation for this research was to answer the following question: Do code ensembles exist which can simultaneously achieve the entire Slepian-Wolf region under message passing decoding? This question still remains unanswered. Although classical LDPC codes can have a large ACPR, it is still not clear if they can be universal or if they are strictly sub-universal. There also seems to be a complexity-universality trade-off at the decoder, which needs to be addressed.

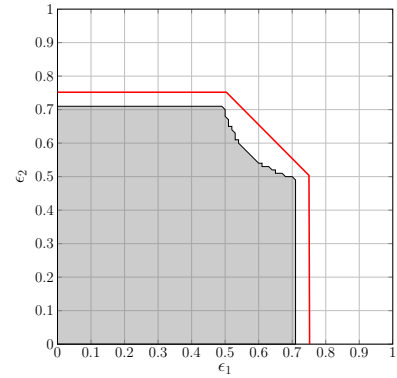

Figure 5. ACPR (Density Evolution threshold) of an optimized (erasure channel) LDPC Code of rate 0.3308